\documentclass[showpacs,preprintnumbers,amssymb]{revtex4-1}%

\usepackage[pdf]{pstricks}
\usepackage{graphicx}
\usepackage{epstopdf}
\usepackage[intlimits]{amsmath}

\usepackage{dcolumn}
\usepackage{bm}

\usepackage[sort&compress]{natbib}
\usepackage[mathscr]{eucal}
\usepackage{color}

\newcommand{\beq}{\begin{equation}}
\newcommand{\eeq}{\end{equation}}
\newcommand{\bey}{\begin{eqnarray}}
\newcommand{\eey}{\end{eqnarray}}
\newcommand{\etal}{{\it et al.}~}

\newcommand{\Rex}[1]{Re_{\scriptscriptstyle{#1}}}
\newcommand{\Rox}[1]{Ro_{\scriptscriptstyle{#1}}}
\newcommand{\Ros}{Ro_{\scriptscriptstyle{S}}}

\newcommand{\xzavg}[1]{\langle #1 \rangle_{\scriptscriptstyle{xz}}}
\newcommand{\xztavg}[1]{\langle #1 \rangle_{\scriptscriptstyle{xz,t}}}
\newcommand{\Xztavg}[1]{\big\langle #1 \big\rangle_{\scriptscriptstyle{xz,t}}}

\newcommand{\half}{\frac{1}{2}}
\newcommand{\uwall}{U_{\scriptscriptstyle{\text{\textsc{w}}}}}

\newcommand{\diss}{\varepsilon}

\newcommand{\Nux}{Nu_{\scriptscriptstyle{\text{max}}}}
\newcommand{\uu}{\xztavg{u^2}}
\newcommand{\vv}{\xztavg{v^2}}
\newcommand{\ww}{\xztavg{w^2}}

\begin{document}
\title{Turbulent states in plane Couette flow with rotation}

\author{Matthew Salewski}
\email{matthew.salewski@physik.uni-marburg.de}
\author{Bruno Eckhardt}
\email{bruno.eckhardt@physik.uni-marburg.de}
\affiliation{Fachbereich Physik, Philipps-Universit\"at Marburg, \\ Renthof 6, D-35032 Marburg, Germany}
\affiliation{J.M. Burgerscentrum, Delft University of Technology, Mekelweg 2, 2628 CD Delft, The Netherlands}

\date{\today}

\begin{abstract}
Shearing and rotational forces in fluids can significantly alter the transport of momentum.
A numerical investigation was undertaken to study the role of these forces using plane Couette flow subject 
to rotation about an axis perpendicular to both wall-normal and streamwise directions. 
Using a set of progressively higher Reynolds numbers up to $Re=5200$, we find that 
the torque for a given $Re$ is a non-monotonic function of rotation number, $Ro$. 
For low-to-moderate turbulent Reynolds numbers we find a maximum  
that is associated with flow fields that are dominated by downstream vortices and calculations of 2-d 
vortices capture the maximum also quantitatively. 
For higher shear Reynolds numbers a second stronger maximum emerges at smaller rotation numbers, 
closer to non-rotating plane Couette flow. It is carried by flows with a markedly 3-d structure and cannot
be captured by 2-d vortex studies. As the Reynolds number increases, this maximum becomes stronger and 
eventually overtakes the one associated with the 2-d flow state. 
\end{abstract}


\maketitle

\section{Introduction}
The effects of rotation on shear flows are relevant in many industrial, geophysical, atmospheric, and 
astrophysical settings. Perhaps the simplest laboratory realization is rotating plane Couette (RPC) flow,
experimentally realized with a plane Couette system on a rotating
table \cite{tillmark1996experiments,FLM_7464588}.
Another closely related system is Taylor-Couette (TC) flow between independently rotating cylinders \cite{couetteetudes,taylor1936fluid}. Both systems
show a variety of different flow states as a function of the two control parameters, related to the shear and the
rotation. The TC system has a third control parameter, the radius ratio, and when this radius ratio
approaches 1, RPC flow emerges out of the TC system.
In addition to the numerous flow states found at low and intermediate shear rates, TC has attracted much attention
with the identification of maxima in torque for varying rotation rates at fixed, high shear rates 
\cite{ PhysRevLett_106_024501, PhysRevLett_106_024502}. 
These studies have also shifted the focus from studies of the different bifurcations and flow patterns 
back to the key physical properties, namely the variation of the total torque required to keep the cylinders in motion with
the external control parameters \cite{wendt1933turbulente,taylor1936fluid,smith1982turbulent,lathrop1992turbulent,lathrop1992transition,lewis1999velocity,ravelet2010influence}.
Several studies \cite{eckhardt2000scaling,dubrulle2002momentum,racina2006specific,eckhardt2007torque,eckhardt2007fluxes} 
of angular momentum transport have sought to explain the observed scalings and to include them in a 
wider context in the study of turbulence through the association of heat transport in the 
Rayleigh-B\'enard system \cite{Physics.5.4}. 

The maximum in the torque seems to occur at a fixed rotation rate, independent of the shear, for high shear rates,
but it seems to have a weak rotation dependence for lower shear \cite{brauckmann2012direct,ostilla2013optimal}. 
The existence and location of the maxima is a consequence of the curvature and the linear stability properties of the flow
\cite{FLM:8684408,brauckmann2013intermittent,merbold2013torque}. In the theoretical explanations, 
the azimuthally-aligned vortices, which occur as the supercritical bifurcation from the laminar baseflow \cite{chossat1985primary}, play a significant role. The flow, in analogy to the corresponding TC-state commonly 
called `Taylor-vortex flow' (TVF), is two--dimensional, and the parallel, counter--rotating vortices advect (relatively) faster--moving fluid from the wall--regions and transport it across to the opposite wall, thereby carrying angular momentum. 
Such vortices will also play a role in RPC flow.

The cylindrical geometry of the TC-system necessarily induces centrifugal instabilities and hence strongly influences
the transport of angular momentum. The curvature is captured by the radius ratio, the ratio between the radii of the
inner and outer cylinders. As this ratio approaches one, TC flow becomes RPC flow, and the transition mechanism changes from centrifugally driven to shear driven \cite{faisst2000transition}. The dependence of the torque on the ratio between 
the radii of the inner and outer cylinders is at the focus of several ongoing studies and the results we present here
provide the limiting behaviour as the radius ratio approaches one.

The numerical simulations which we present here cover a wide range of rotation parameters for a small set of shear
parameters, since we are mainly interested in the way rotation influences flow properties and torque values. To see 
this most clearly, we study the flows for fixed shear rates and varying rotation rates.

The outline of the paper is  as follows. In section II we discuss the system, the momentum transport and the
numerical aspects. In section III, we then discuss our results, first for the momentum transport and then for the mean profiles.
We conclude with a few remarks in section IV.

\section{Rotating plane Couette flow and its numerical simulation}

\subsection{System and Parameters}
\begin{figure}[h]
\includegraphics[scale=1.]{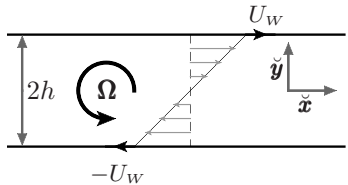}
\caption[]{The orientation of the plane Couette system with wall velocity $\uwall$ and an imposed global rotation, $\mathbf{\Omega}$. The orientation of the rotation vector is out-of-the-page and shows the anticyclonic rotation which this paper mainly deals with.}\label{Figure_1}
\end{figure}

In the rotating frame, the Navier-Stokes equations for an incompressible fluid in planar geometry are
\bey\label{RotNSE}
\partial_t \pmb{U} + \big(\pmb{U}\cdot\nabla\big)\pmb{U}-2\pmb{U}\times\mathbf{\Omega} & = &-\frac{1}{\rho}\nabla\Pi+\nu\Delta\pmb{U},\\ 
\nabla\cdot\pmb{U} & = & 0;
\eey
the pressure is modified by the inclusion of a centrifugal force,
\beq
\Pi(\pmb{x},t) = P(\pmb{x},t)-\frac{\rho}{2}(\mathbf{\Omega}\times\pmb{x})^2.
\eeq
We use $(x,y,z)$ to denote the streamwise, wall-normal, and spanwise directions, respectively, and similarly use the notation, $\pmb{U}(\pmb{x},t)=(\uwall y + u(\pmb{x},t),v(\pmb{x},t),w(\pmb{x},t))$, for the velocity components in these directions. The top and bottom walls are separated by a distance $2h$ and move oppositely with a velocity difference of $2\uwall$. The domain is periodic in the stream- and spanwise directions. The rotation occurs around a spanwise axis, hence, $\mathbf{\Omega}=\Omega\mathbf{e}_{z}$. 
In fig.~\ref{Figure_1}, we show the geometry of this system, the orientation of the axes and walls, and the applied rotational force.

The equations are non-dimensionalized using the wall-velocity, $\uwall$, and half of the channel height, $h$; this establishes the shear Reynolds number, 
$Re$, as well as the rotation parameter, $\Ros$,
\begin{eqnarray}
Re=(\uwall h)/\nu\\
\Ros=(2\Omega h)/\uwall \,.
\end{eqnarray} 
The Rossby number used, e.g., in geophysics, is the inverse of $\Ros$.
Viscosity and the intrinsic lengthscale $h$ then define a velocity scale $v_{\scriptscriptstyle{\nu}} = \nu/h$, with which
one can construct a `viscous' rotation number, $\Rox{V}=2\Omega h^2/\nu$, 
which is the ratio of Coriolis to viscous forces; it is also
known as the (inverse) Ekman number \cite{greenspan1968theory}.
The two rotation numbers are connected through the relation
\beq
\Rox{V}=Re\Ros.
\eeq

The orientation of the rotation vector significantly affects the flow. The vorticity of the laminar baseflow, $\nabla\times\pmb{U}_{\!\scriptscriptstyle{\mathrm{b}}}(y)=-\uwall\mathbf{e}_{z}$, is constant and negative in the spanwise direction. When the vorticity of the laminar profile and the rotation vector are parallel, the flow is known as `cyclonic' and the rotation has a stabilizing effect on the flow; for anti-parallel pairing of these vectors, the `anti-cyclonic' effect destabilizes the flow. 
The differences between cyclonic and anti-cyclonic rotation can be understood by considering the Coriolis' role in the equations of motion, $\pmb{U}\!\times\mathbf{\Omega}$.
In the cyclonic orientation, the high--speed fluid near the walls is redirected towards the walls, where viscous dissipation is at its largest; perturbations here would be damped relatively quickly compared with their ability to elicit a (transient) response. This stabilization of the flow preserves the linear stability of non-rotating plane Couette flow \cite{schmid2001stability}. Anti-cyclonic rotation causes the high--speed fluid near the walls to be turned to the flow's interior, setting up an instability
that resolves itself through the formation of the counter-rotating vortices aligned in the streamwise direction.
They correspond go the TVF in TC and will hence be referred to by this label.
The vortices are invariant in the downstream direction (azimuthal in the TC geometry), so 
that the flow is effectively two-dimensional.

There are further connections between RPC flow and TC flow.
In both systems, the base flow is unidirectional, parallel to the walls, and has a (relatively) simple dependence on the distance from the walls. Furthermore, the rotation is around a vector that is orthogonal to both the base flow direction and wall-normal direction; this is the $z$-axis in RPC, making the spanwise direction the axial direction in TC. The 
streamwise direction then translates to the azimuthal direction. The link between these systems was studied by
Dubrulle \etal\cite{dubrulle_095103} who sought a set of parameters that could be used in rotating shear flows 
in both cylindrical and planar geometries. 
Their parameters $\Rex{D}$ and $\Rox{D}$ depend on the radius ratio of the TC system, 
but are chosen such that they remain finite in the limit of the planar system. Their parameters (marked with 
a subscript "D") are related to ours  by
\bey
 Re &=& \Rex{D}/4\label{DubrulleRe}\\
\Ros&=&-\Rox{D}\label{DubrulleRo}.
\eey
In addition,  the stability properties of RPC are closely related to TC \cite{lezius1976roll}. Both systems have transitions to turbulence via supercritical and subcritical mechanisms depending on the parameter values. In both cases there is a supercritical bifurcation with the laminar baseflow transitioning to two--dimensional TVF \cite{taylor1923stability}, with vortices aligned in the streamwise direction. Bifurcations from the Taylor vortices to the wavy  vortices \cite{chossat1985primary,iooss1986secondary,nagata1988wavy} and their variations \cite{nagata1997three,nagata1998tertiary}, lead to turbulence via the Ruelle-Takens scenario \cite{guckenheimer1983strange}.
On the subcritical side, the progress towards turbulence is via the onset of growing turbulent spots, becoming turbulent stripes which eventually fill the entire domain \cite{manneville1980different,lundbladh1991direct,duguet2010formation}; 
this latter view has  seen recent confirmation in pipe flow \cite{avila2011onset}.

The dynamics of both systems remain similar even beyond the bifurcation diagrams. E.g., the experimental studies in RPC \cite{tillmark1996experiments,alfredsson2005instability,hiwatashi2007experimental,FLM_7464588} show a strong overlap with the well-known results from TC \cite{coles1965transition,andereck1986flow}. Tsukuhara \etal \cite{FLM_7464588} 
made a detailed parameter scan of the dynamics seen in RPC, and created a state space diagram in the spirit of 
Andereck \etal \cite{andereck1986flow}, where one can directly see an overlap in the dynamics. However, we note that the
parameters of Andereck \etal \cite{andereck1986flow} are
the outer and inner Reynolds numbers, $\Rex{o}$ and $\Rex{i}$ and hence are different from $\Rox{V}$ and $Re$
used in Tsukuhara \etal \cite{FLM_7464588}. 
The relation between these sets of parameters is given by Dubrulle \etal \cite{dubrulle_095103}.

\subsection{Force and Momentum Current}
The physical quantity we use to distinguish different flow states is the force or momentum
flux between the moving walls. 
Here, we briefly derive this quantity and relate it to its analogue in the TC--system. Following Eckhardt, \etal\cite{eckhardt2007torque}, we start with the streamwise component of the velocity field,
decomposed into a base-profile and the fluctuations as $U = \uwall y + u$,
\beq\label{baseflow}
\partial_t u = -v \uwall -\uwall\partial_x u- \big(\pmb{u}\cdot\nabla\big)u - \partial_x p + \nu\Delta U + 2\Omega v. 
\eeq
${\bf u} = (u,v,w)$ is the vector of fluctuating components.
Though it vanishes in the dissipative term of the above equation, we keep $\uwall$ as it is retained in the momentum 
current (see below). We define spatial averages in the stream- and spanwise directions by
\beq\label{define_xztavg}
\xzavg{ A(y,t) } = \frac{1}{L_x L_z}\int_0^{\:L_x}\!\!\!\int_0^{\;\;L_z} \!\!\! A(x,y,z;t) \, dx\,dz.
\eeq
and sometimes also time-averages over the statistically stationary states.
Then equation (\label{baseflow}) becomes equivalent to conservation of the momentum current
\beq\label{mmtm_current1}
J^u = \xztavg{u v} - \nu\partial_y \xztavg{U}.
\eeq
Near the walls it reduces to the local `wall shear stress' $\tau_w$ \cite{pope2000turbulent}.
To make the connection of the present system to the torque measurements and dimensionless
quantities in the TC system, we divide the 
momentum current by its laminar value to obtain a dimensionless momentum flux, $Nu$, 
which serves as the analogue of the Nusselt number in TC flow and Rayleigh-B\'enard convection; see \cite{eckhardt2000scaling,eckhardt2007fluxes}. 
Noting that ${J}^u_{\scriptscriptstyle{\text{lam}}}=\nu \uwall/h$, the force-Nusselt number is defined as 
\beq\label{mmtm_current_Nu}
Nu= \frac{J^u}{{J}^u_{\scriptscriptstyle{\text{lam}}}} = \frac{h}{\uwall \nu}\Big(\xztavg{u v} - \nu\partial_y \xztavg{U}\Big).
\eeq
As in other cases, it consists of  a Reynolds-stress, $\xztavg{u v}$, and a viscous gradient of the 
streamwise velocity in the wall-normal direction, and while each of these depends upon $y$, the 
momentum flux, spatio-temporally averaged, is $y$--independent  (see \cite{eckhardt2007torque} or \S7.1 in \cite{pope2000turbulent}). 
Since $J^u$ is constant in $y$, its value may be taken at the boundaries of the system, $y = \pm 1$. 
For rigid, impermeable walls, $ \xztavg{u v}\big|_{\scriptscriptstyle{y=\pm 1}} = 0$ since 
$v\big|_{\scriptscriptstyle{y=\pm 1}} = 0$,
hence
\beq\label{mmtm_current2}
J^u \big|_{\scriptscriptstyle{y=\pm 1}} =  - \nu\partial_y \xztavg{U}\big|_{\scriptscriptstyle{y=\pm 1}}.
\eeq
The friction Reynolds number $\Rex{\tau}$ is defined with the wall shear stress  $\tau_{\scriptscriptstyle{W}}$ as
\beq\label{fricRe1}
\Rex{\tau} \equiv  \frac{h}{\nu}\sqrt{\frac{\tau_{\scriptscriptstyle{W}}}{\rho}} = \frac{h}{\nu}\sqrt{J^u}.
\eeq
so that the relation between Nusselt number and $\Rex{\tau}$ becomes
\beq\label{fricRe2}
\Rex{\tau} = \sqrt{Re Nu}.
\eeq
One can similarly construct a measure of the skin-friction coefficient,
\beq\label{fricfac1}
c_f = \frac{2\tau_{\scriptscriptstyle{W}}}{\rho \uwall^2},
\eeq
which can be found in shear flows near a wall \cite{pope2000turbulent,landau1987fluid}.
The skin-friction is related to the Nusselt number analogue 
in TC by \cite{lathrop1992transition,ravelet2010influence},
\beq\label{fricfac2}
c_f = \frac{2}{\uwall^2}\bigg(\frac{\uwall \nu}{h}Nu\bigg) = \frac{2}{Re}Nu.
\eeq
A second point to note is that $J^u$ does not contain the rotation rate explicitly, since
the spatial average of the incompressibility condition leads to $\xzavg{v}=0$. 
As a consequence, the effects of the rotation have to show up in the momentum 
transport through their effects on the flow and the Reynolds stresses and gradients at the walls.

\subsection{Numerical aspects}
For the direct numerical simulations (DNS) we use the code \verb|channelflow| developed by 
J. F. Gibson \cite{channelflowcode}, used and verified extensively, see e.g. 
\cite{schneider2008laminar,gibson2008visualizing,gibson2009equilibrium,halcrow2009heteroclinic,schneider2010snakes,schneider2010localized,kreilos2012periodic}. 
For this work, we implemented an OpenMP interface, extending \verb|channelflow| to run on shared-memory processors. 
We also configured the FFTW3 library \cite{frigo2005fftw} that \verb|channelflow| uses to run threaded, 
obtaining a moderate increase in speed. 

To treat Coriolis forces we used the so-called `rotational' form \cite{canuto1988spectral} of the nonlinear 
term, $\pmb{U}\cdot\nabla \pmb{U}=\half\nabla \pmb{U}^2-\pmb{U}\times\nabla\times\pmb{U}$,
so that Eq.~\ref{RotNSE} becomes
\beq
\partial_t \pmb{u} - \pmb{U}\times\big((\nabla \times\pmb{U}) + \Ros\pmb{e}_z\big) = -\nabla\big(p + \frac{1}{2}\pmb{U}^2\big)+\frac{1}{Re}\Delta\pmb{u}.
\eeq
The code evolves the full velocity field ${\pmb U}=
y \mathbf{e}_x+{\pmb u}$, containing both the laminar profile and the fluctuations, in the nonlinear term.
The time-stepping for \verb|channelflow| is a semi-implicit, multistep--backwards finite--difference scheme with the modified nonlinear term including the Coriolis force, treated explicitly while the solution of the linear terms is done implicitly; this scheme is a common treatment for including the Coriolis force, see for example \cite{rincon2007self,brethouwer2012turbulent}.

The rotational form of the nonlinear term requires dealiasing \cite{zang1991rotation}, and hence we use 
a $2/3$--dealiasing rule for all of our simulations. The gridpoint resolutions for our simulations are 
$N_x\times N_y\times N_z=108\times71\times108$, $192\times81\times192$, $256\times113\times256$, $384\times129\times384$ for $Re=650$, $1300$, $2600$, and $5200$, respectively. 
Resolution was checked by statistical convergence of volume-averaged quantities for successively increased resolutions, specifically using the relationship that $\nu J^u/h^2 = \diss$ as suggested in \cite{ahlers2009heat,shishkina2010boundary} for RB and \cite{eckhardt2007torque,brauckmann2012direct} for TC.

We examine turbulent regimes for $Re=650$, $1300$, $2600$,  and $5200$. The Reynolds number of $Re=1300$, and multiples thereof, was chosen as this has been used a number of studies for plane Couette flow, with and without rotation \cite{kristoffersen1993numerical,bech1995investigation,bech1996secondary,bech1997turbulent,barri2010computer}. Preliminary tests using $Re=1300$ showed that box size did not have an influence on the critical rotation number when the width of the box was above $0.5\pi$, and similarly for the streamwise length. We decided then to use boxes of $L_x\times L_z = 4\pi h\times 2\pi h$. Even though this box is smaller than the ones used in other study, we find that 
the friction Reynolds numbers $\Rex{\tau}$ match other simulations in larger domains rather well \cite{papavassiliou1997interpretation,debusschere2004turbulent,tsukahara2006dns,brethouwer2012turbulent}
The components of $J^u$ given in \cite{barri2010computer} for $\Ros=0$, and  $0.7$ agree with our results.
Our results are also consistent with other large-domain studies but with different values of $Re$, though mostly without rotation, \cite{papavassiliou1997interpretation,debusschere2004turbulent,tsukahara2006dns,brethouwer2012turbulent}.
Lastly, recent numerical studies of the turbulent TC--system with short azimuthal (streamwise) lengths are, in some cases, consistent with available experimental measures \cite{brauckmann2012direct} and in others agree solidly \cite{merbold2013torque}.

In addition to the turbulent simulations, \verb|channelflow| is capable of finding and continuing exact coherent solutions, such as Taylor vortices, using a Newton-hookstep search-algorithm \cite{viswanath2007recurrent} with pseudo-arclength continuation scheme. Since the search-algorithm employs the DNS for the Newton method, no additional modifications were needed to include the Coriolis forcing. The continuation program was adapted to follow solutions in rotation number, but this required no significant changes to the main algorithm.

\section{Results}
We begin the presentation of our results with the dependence of the friction force on shear and rotation, followed
by a discussion of the velocity fields and the mean profiles.

\subsection{Global force measurements}
\begin{figure}[h]
\includegraphics[scale=0.5]{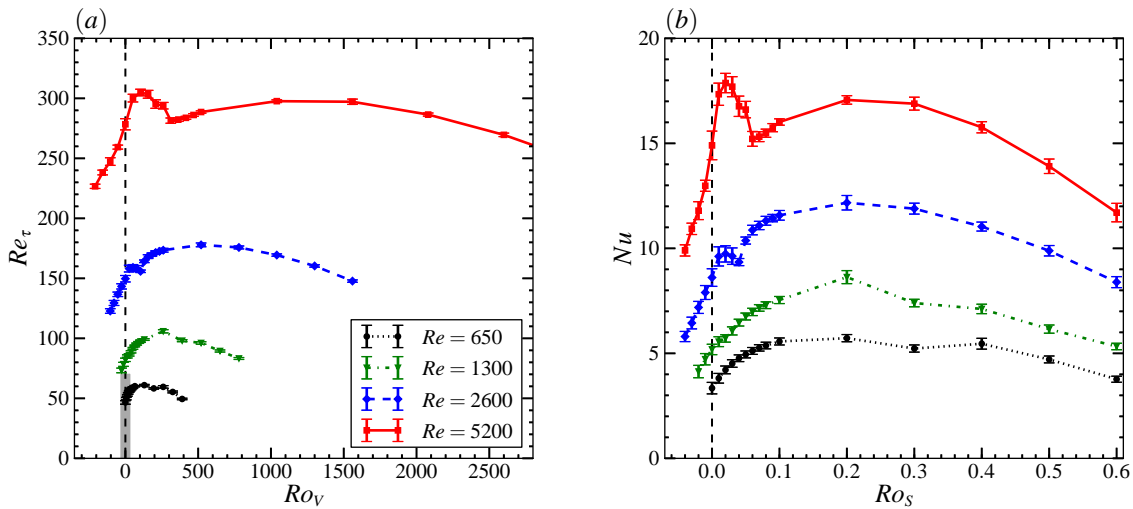}
\caption[]{The friction Reynolds number, $\Rex{\tau}$, and momentum transport current, $Nu$, 
against the rotation parameters $\Rox{V}$ and $\Ros$, 
respectively. 
The small shaded region near the base of the $\Rex{\tau}$-axis in $(a)$ gives the horizontal 
extent of the domain described in \cite{FLM_7464588}. The error bars indicate the statistical 
uncertainty extracted from the fluctuating signals.}
\label{Figure_2}
\end{figure}

The results for the wall-normal momentum flux are shown in 
fig.~\ref{Figure_2}. One frame shows the friction Reynolds number, $\Rex{\tau}$, vs. $\Rox{V}$, the other
the momentum-Nusselt number against $\Ros$. 
For all shear Reynolds numbers shown, the general trend is that the torque increases for increasing anti--cyclonic
rotation and reaches a maximum value in the low-to-moderate anti-cyclonic rotation regime. 
For low shear Reynolds numbers there is a single maximum, but for higher $\Rex{S}$ there are two: a narrow one at small rotation and a broad one for larger $\Ros$.
Moreover, we found that rotation suppresses the turbulence, as noted before in \cite{bech1996secondary,bech1997turbulent,barri2010computer}. 
One notes that the maxima in the $\Rex{\tau}$ vs. $\Rox{V}$ plot
are not aligned and the range that can be covered varies with $\Rex{\tau}$. In contrast, when
plotted against $\Ros$ the maxima do line up, suggesting that $\Ros$ is the more appropriate parameter. 

The cyclonic regime in RPC is entirely subcritical, bifurcating from the laminar baseflow only when $Re$ goes to infinity.
The empirically-found state-space of \cite{FLM_7464588} shows sustained turbulence for this region. 
Our domain is considerably reduced in comparison and has an increased likelihood for decay. For this reason, 
the cyclonic rotation numbers we report here are chosen from simulations which sustain the turbulence long enough 
to obtain reasonable statistics ($\simeq 5000$ nondimensional time units).

\begin{figure}[h]
\includegraphics[scale=0.5]{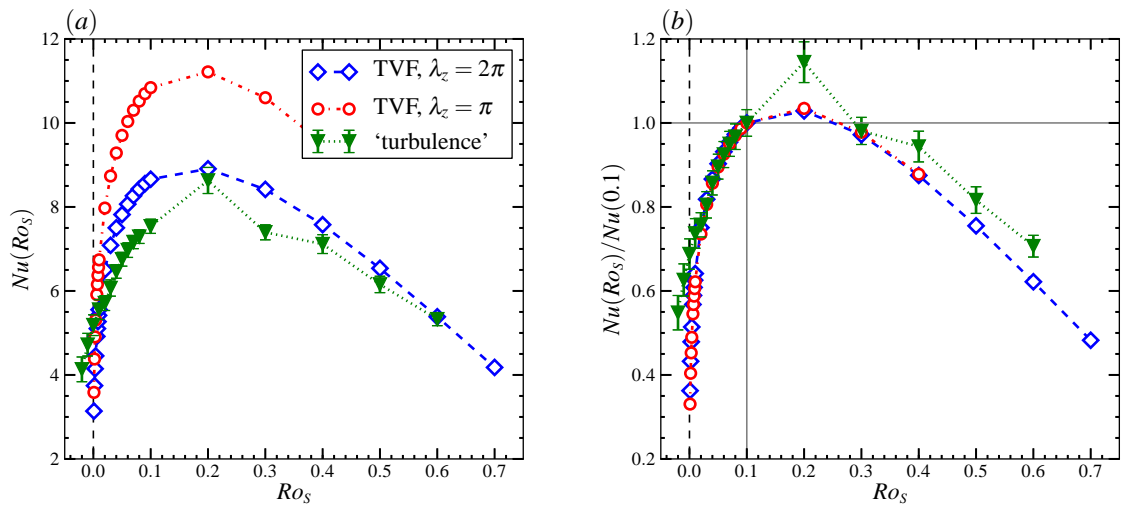}
\caption[]{Comparison of the momentum transport current, $Nu$, again as a function of $\Ros$, for a fully turbulent simulation and the exact Taylor vortex flow solutions at $Re=1300$; graphs $(a)$ and $(b)$ show the absolute and rescaled value of $Nu$, respectively. In both graphs, the shape of the turbulent momentum transport curve resembles those of the computed TVF solutions.}\label{Figure_3}
\end{figure}

The global shape of the torque variation with $\Ros$ has an unexpected explanation: 
it follows closely the curve of the torque $Nu_{\scriptscriptstyle{\mathrm{TVF}}}$ of 2-d, longitudinally
aligned vortices.
This is demonstrated in fig.~\ref{Figure_3}, where we compare the data from the turbulent simulation for $Re=1300$ with the 
momentum transport for two 2--$d$ Taylor vortex solutions, one with a spanwise wavelength coinciding 
with the boxwidth, $\lambda_z=2\pi h$, and the other with half the wavelength; these solutions correspond 
to a flow with one and two vortex pairs, respectively. 
They were found using the linear instability of the laminar flow at  a low Reynolds numbers and 
one rotation number, usually $\Ros=0.1$, and were then continued to higher Reynolds and different rotation
numbers using the hookstep Newton-solver implemented in channelflow \cite{gibson2009equilibrium}. 
With this procedure it is a  simple matter to reduce the streamwise box-length, taking advantage of the 
solution's two--dimensionality, to $1/8$ of the original size.

Both 2-d vortex states in fig.~\ref{Figure_3} show similar variations with $\Ros$, and the
one vortex state also agrees quantitatively with the turbulent RPC flow simulation. That the TVF--solution dominates the
momentum transport flux can be rationalized by the presence of the streamwise vortices and the redirection of energy 
into the wall-normal components that has a significant impact on the transfer of momentum \cite{rincon2007self}. The 
role of the noise generated by the turbulent fluctuations on streamwise vortices reduces their transport effectiveness, 
hence the discrepancy between the turbulent and exact solutions. 

\begin{figure}[h]
\includegraphics[scale=0.5]{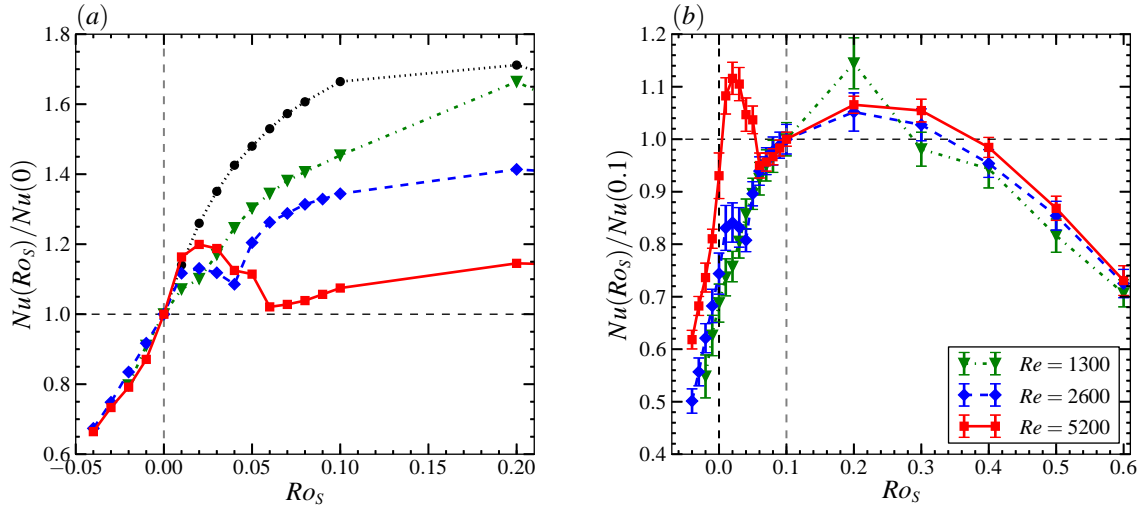}
\caption{The momentum transport as in fig.~\ref{Figure_3}, normalized their respective values at zero-rotation for $(a)$ and 
$\Ros=0.1$ for $(b)$. Here, graph $(a)$ shows the development of the second, `narrow' peak (see text for details) while 
panel $(b)$ shows that the original peak corresponds to the Taylor vortex flow peak.}
\label{Figure_4}
\end{figure}

Returning to fig.~\ref{Figure_2}, the most striking feature is the presence of a second peak at 
$\Rox{V}\sim0.02$ for $Re\geq2600$. As quantitative measures of the increased momentum transport, 
we normalized the momentum transport flux by its non-rotating value, fig.~\ref{Figure_4}$(a)$, and using a value 
nearer to the TVF--maximum, fig.~\ref{Figure_4}$(b)$. 
Fig.~\ref{Figure_4}$(a)$ highlights the relative increase in the region for smaller rotation numbers, $\Ros\lesssim0.1$. 
It can be seen from this figure that, in general, the Taylor vortices become weaker when the Reynolds number is 
increased. For $Re=650$, there is a 77\% enhancement in momentum transport compared to the non-rotating case; this drops to 54\%  for $Re=1300$, and is reduced further to 41\% and 15\% for Reynolds numbers 2600 and 5200, respectively. 
Conversely, we see that the second peak, henceforth referred to as the `inner peak' as it is closer to the $\Ros=0$--axis, increases with Reynolds number. In the curve for $Re=1300$, there is a slight bump at $\Ros = 0.02$; it will be demonstrated later that this corresponds to the same peak in the larger $Re$--cases. Given that the steps taken in rotation number, $\Delta\Ros = 0.01$ for $\Ros < 0.1$, do not necessarily coincide with the actual positions of the maxima, we 
estimate enhancements of 10\%, 13\%, and 20\% for $Re=1300$, $2600$, and $5200$, respectively. 
The value of $Nu(0)$ also increases with Reynolds number as in fig.~\ref{Figure_2}, but not as rapidly as the new peak.
The accompanying plot, fig.~\ref{Figure_4}$(b)$, shows again the coincidence of the curves when using a normalizing value taken from the TVF--dominated flows. The large deviation of the $Re=1300$ curve at $\Rox{S}=0.2$ is due to this flow having two vortex pairs (see below).

The respective $\Ros$-position of the inner peak does not change appreciably with Reynolds number over the 
range studied here. This is in contrast to the TVF-peak, which within the resolution in $\Ros$ available seems to 
move towards slightly higher $\Ros$ with increasing Reynolds number, as seen in fig.~\ref{Figure_4}$(b)$.

\subsection{Velocity fields}

\begin{figure}[h]
\includegraphics[scale=0.5]{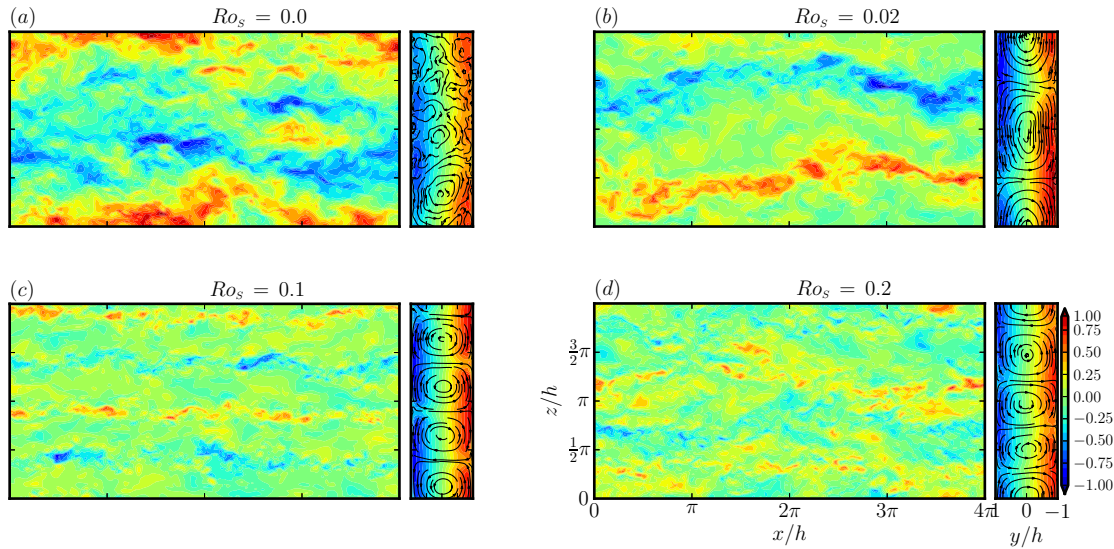}
\caption{Images of the streamwise flowfield, $u$, in the $xz$--plane at $y=0$ (left) along with its streamwise-averaged flowfield, $\big\langle \pmb{u}\big\rangle_x $, shown as an $yz$--plane at $x=0$ (right) for $Re=5200$. The velocity in the $xz$--plane has been rescaled by a factor of $2.5$ in comparison to the color-bar scale; this was done to increase the contrast in the flow. The streamlines are used here to provide a qualitative impression of the flowfield.
}\label{Figure_5}
\end{figure}

While the outer, broad maximum is connected with 2-d Taylor vortices, the inner peak has a fundamentally different 
dynamics. To demonstrate this, we show the flow fields and their properties for various cases of rotation, 
$\Ros=0$,  $0.02$, $0.1$, and  $0.2$, in fig.~\ref{Figure_5}. 
The case of the highest Reynolds number, $\Rex{s}=5200$ shows the most notable differences. 
The figure shows snapshots of the streamwise velocity field plotted in the $xz$--plane in the middle of the gap, 
$y=0$, 
and its streamwise averaged profile in the $yz$--plane with the averaged wall- and spanwise flows depicted 
with streamlines (the local density of parallel streamlines indicates the strength of the flow in this direction).

The first such case we present is that for zero-rotation in fig.~\ref{Figure_5}$(a)$. 
For the non-rotating case, there is no Coriolis force to maintain large streamwise structures, such as Taylor vortices. 
Some large streamwise streaks do appear, but these do not persist for appreciable times.
The accompanying vortices, whose normal extent does not fill the channel's height, are similarly fleeting. The streamwise average of the flowfield 
shows some of the remaining cross-stream flows, revealing little coherence and a interfering tangle of vortices. Without the anti-cyclonic rotation to sustain the vortices for extended periods of time, the wall-wise transport of momentum can only be less efficient without the collaboration among the present structures.

When rotation is included, even if for small values like $\Ros\simeq0.01$, we see a dramatic change in the flow state. The image in fig.~\ref{Figure_5}$(b)$ corresponds to the inner peak for $Re=5200$. In comparison to the non-rotating case, we see that there are significant vortical structures spanning the streamwise length; they are contiguous for all times observed. Together, the vortices are also flattened out and fill the $yz$--plane, in both average and instantaneous (not shown) flows. It seems to be a Taylor vortex pair, however, with a streamwise modulation, much like that of the so-called wavy-vortex flow which is the first bifurcation from the TVF--solution in both RPC \cite{nagata1988wavy} and Taylor-Couette flow \cite{chossat1985primary,iooss1986secondary}. Observing movies of this flow state shows that they are not constant in time, with the modulation increasing and decreasing in amplitude. This was also observed by Komminaho, \etal \cite{komminaho1996very} but in non-rotating plane Couette flow; we note a marked distinction between this and the non-rotating flowstate, mainly in the spatial coherence. 

Finally, we show flowfields near to the second peak in fig.~\ref{Figure_5}$(c)$ and fig.~\ref{Figure_5}$(d)$, for $\Ros=0.1$ and
$0.2$, respectively, leading to the peak within the range $0.1\leq\Rox{S,\mathrm{max}}\leq0.3$. We see in both images that the flow is mostly organized as two--dimensional, with some small-scale fluctuations; the overhead, mid-plane plot (left in both figures) shows two pairs of (alternating) by high-speed streaks and the $xz$--averaged plot (right) distinctly shows two vortex pairs. Observing these flowfields in time shows that the streaks are being rapidly advected in the streamwise direction. The main occurence in this rotation number region is the strengthening of the vortices, matched by increased advection of the streaks. Both of these features are consistent with the laminar case of TVF \cite{rincon2007self}; it is as if the turbulent fluctuations do not matter.

For $\Ros=0.1$, the streaks are relatively contiguous in the streamwise direction and there seem to be no large-scale fluctuations to upset the two-dimensionality of the flow. One can appreciate this in the lefthand image of fig.~\ref{Figure_5}$(c)$. The vortices for this rotation number are always present and relatively robust in their spanwise positions. Similarly, and seen in the righthand image, the vortex-diameters are roughly equal and also stable in time. Films of the simulations were on the order of 300 time-units, and snapshots separated by 50 time-units over the 5000 time-unit simulation length showed little positional changes overall. We note that this region of quasi-$2D$ flow begins, seemingly rather abruptly, for $\Ros\geq0.06$ and continues until the peak.

\begin{figure}[h]
\includegraphics[scale=0.5]{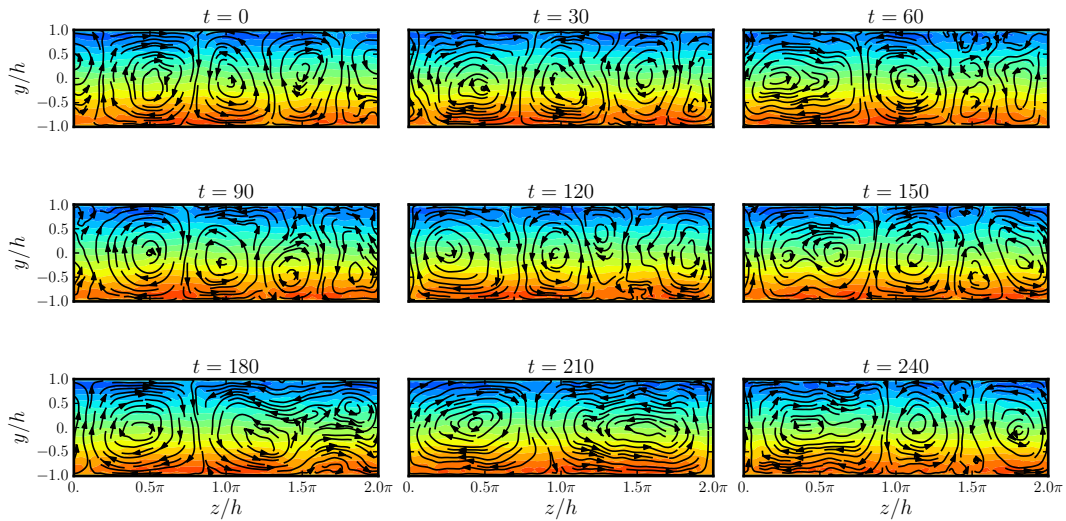}
\caption{Snapshots of the streamwise-averaged flowfield, $\big\langle U_w+u\big\rangle_x $, for $Re=5200$ and $\Ros=0.3$ as it evolves in time, $\Delta t=30$, from a 2--vortex pair state, $t=0$, to a single pair state, $t=210$; the starting time is arbitrary. While the timestep used here misses some transitional events, there is an overall sense of continuity in the large-scale structures.}\label{Figure_6}
\end{figure}

The flow beyond the peak is markedly different. In fig.~\ref{Figure_5}$(d)$, the streaks do not show the same measure of spatiotemporal coherence. An interesting feature of the vortices near the peak is an oscillating spanwise compression/extension fluctuation; an occurrence of this can be seen in the figure where the vortices centered at $0.5\pi$ and $1.5\pi$ are of different sizes. 
Moreover, for $\Ros\leq0.1$, such fluctuations do not occur but when $\Ros\geq0.3$, the fluctuations are significantly larger, and demonstrate a competition between state with one and two vortex pairs, see fig.~\ref{Figure_6}. 

The visualizations give evidence that the peaks found in the momentum flux are driven by the coherent structures 
which are distinguished as being 2-- or 3--dimensional. As a measure for the coherent structures we can take the
wallwise-averaged energy in the spectral modes, $\widetilde{E}_{ln}$, where the indices stand for the
downstream and spanwise Fourier modes. 
With the modulation seen in fig.~\ref{Figure_5} we expect a contribution from the $\widetilde{E}_{10}$-mode for 
the spanwise rolls, and from the $\widetilde{E}_{01}$- and $\widetilde{E}_{02}$-modes for the 
single- and double-vortex pair states, respectively. The relation that gives the spectral energy for a given 
Fourier mode is 
\begin{equation}\label{spectralE}
 \widetilde{E}_{ln}(t) = \frac{1}{2}\sum_{m =0}^{N_y -1} \Big| \widehat{\tilde{\pmb{u}}}_{lmn}(t) \Big|^2, 
\end{equation}
where $\widehat{\tilde{\pmb{u}}}_{lmn}$ are the complex Fourier-Chebyshev coefficients. 
The wallwise-averaged modes considered below are limited to $l,n=1,\ldots,4$; including $l,n \geq 5$ gives no further information as these are comparatively small. Focusing on the orthogonal directions is a choice made empirically; while mixing between modes is found, the full set of $\widetilde{E}_{0n}$ shows low-index modes with either $l=0$ or $n=0$ are usually the strongest in the anti-cyclonic region where quasi-$2D$ flows are encountered.

\begin{figure}[h]
\includegraphics[scale=0.5]{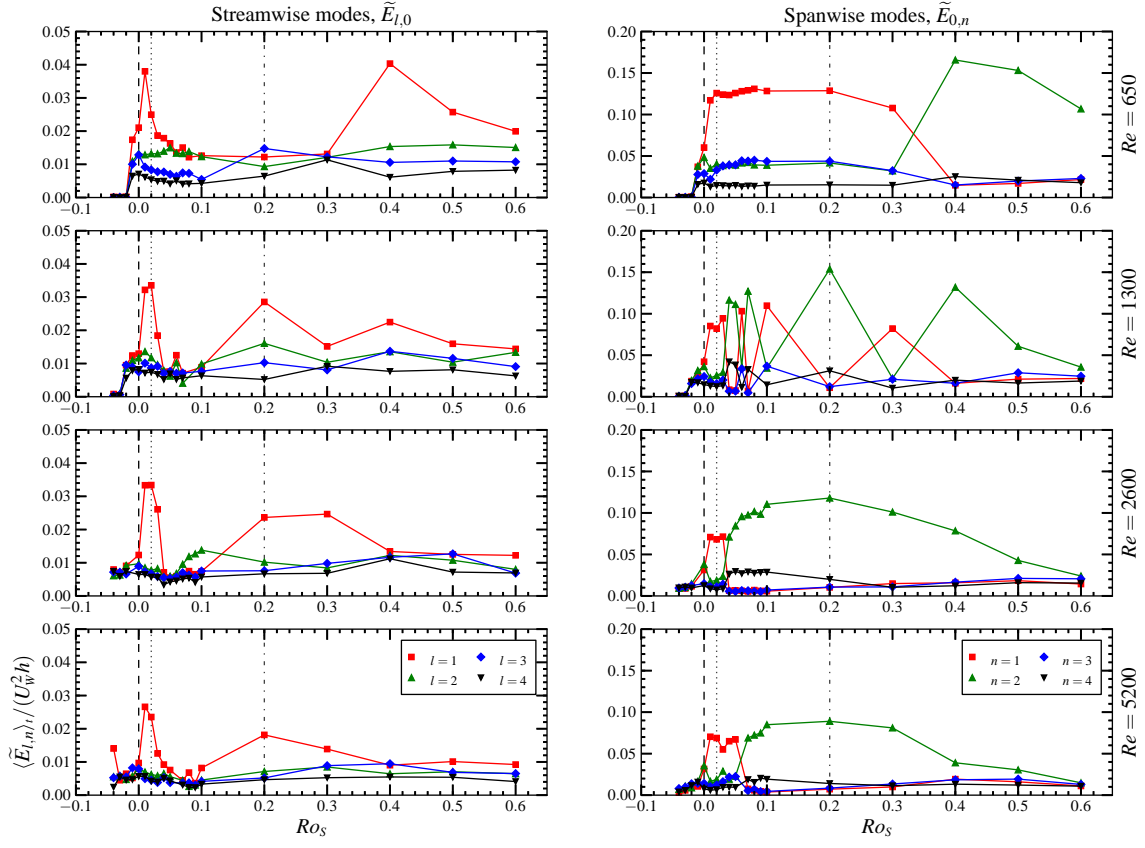}
\caption{Left and right, respectively, are the stream- ($n=0$) and spanwise ($l=0$) spectral energy modes, $\langle\widetilde{E}_{ln}\rangle_{\scriptstyle{t}}$ for mode-numbers $l,n\in[1,4]$, averaged in time, plotted against the rotation number, $\Ros$. Note that the range for the streamwise modes is roughly a factor of four bigger.}\label{Figure_7}
\end{figure}

The plots in fig.~\ref{Figure_7} show that the $\widetilde{E}_{10}$-mode has a strong presence in an acute region of $\Ros$ eventually becoming the narrow peak for larger $Re$, and that this is rapidly reduced for slightly larger rotation numbers where the flow is primarily two--dimensional. The $\widetilde{E}_{10}$-mode re-emerges at the broad peak, 
thereafter all $\widetilde{E}_{l0}$-modes with $l\neq0$ decrease. There are some additionally curious features such as $\widetilde{E}_{01}$-mode being strongest for rotation numbers near to the narrow peak value and that this eventually changes in all cases to the $\widetilde{E}_{20}$-mode being largest; in addition, the onset of the latter mode is delayed when $Re=650$ and fluctuates with $\Ros$ for $Re=1300$. These two observations suggest at least some competition between single-pair and two-pair vortex states, with the latter becoming the more stable of the two as Reynolds number is increased. The wavelength of the two-pair state is closer to the optimal wavelength suggested by numerical simulations of the Taylor-Couette system with a comparable rotation number \cite{brauckmann2012direct}.

\subsection{Mean profiles and fluctuations}

\begin{figure}[h]
\begin{center}
\includegraphics[scale=0.5]{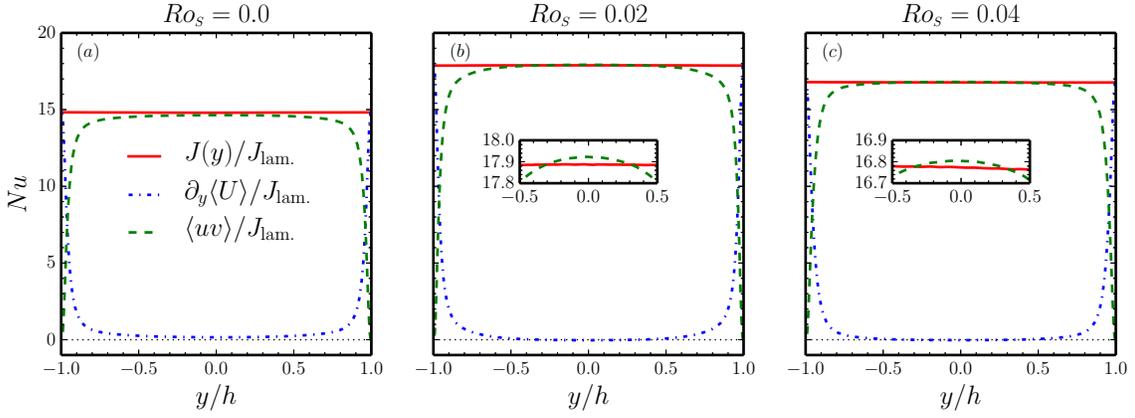}
\caption{The momentum transport and its components' profiles normalized by the laminar momentum transport, $J_{\scriptstyle{\mathrm{lam.}}}$ which gives these quantities in their absolute scale so that they can be compared with fig.~\ref{Figure_2}$(b)$. The insets in graphs $(b)$ and $(c)$ magnify the region near $y=0$ where $\xztavg{uv}$ intersects and becomes larger than $Nu$, indicating a region where $\partial_y\Xztavg{U} < 0$.}\label{Figure_8}
\end{center}
\end{figure}

We now wish to understand the increase in the momentum transport associated with the narrow peak, 
and to contrast it with the broad peak. Firstly, in the definition of $Nu$ in Eq.~\ref{mmtm_current_Nu} there are two 
contributions, $\xztavg{uv}$ and $\partial_y\xztavg{U}$, which vary with the $y$-position. Profiles for ${J}^u$ and 
these components are plotted in fig.~\ref{Figure_8} for $\Ros$ near the narrow peak.

At and around the narrow maximum, seen in graphs $(b)$ and $(c)$, the quasi-Reynolds stress component $\xztavg{uv}$
is larger than $Nu$ for a region centered in the middle of the channel and spanning roughly $40\%$ of the width. Though 
the difference is not large, with $(\xztavg{uv}-Nu)/Nu\lesssim 0.2\%$, it implies that the gradient of the mean-flow, $\partial_y\xztavg{U}$, must be negative in this region. There is no analogous finding for components' profiles corresponding to 
the broad peak (not shown), so that this feature is unique to the narrow peak. 

\begin{figure}[h]
\includegraphics[scale=0.5]{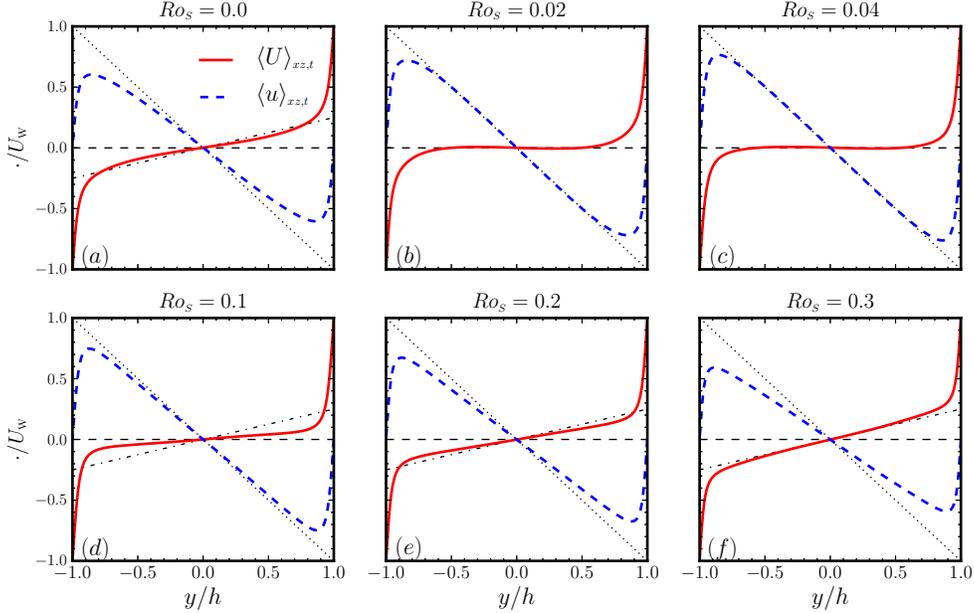}
\caption{The streamwise velocity profiles for the total meanflow, $\xztavg{U}=\uwall y+\xztavg{u}$, and the mean flow 
excluding the laminar contribution, $\xztavg{u}$,  for $Re =  5200$, using various rotation numbers associated with the 
narrow (top row) and broad (bottom row) peaks. Additional lines are added for reference: dotted line with slope 
of $-1$, dashed with slope 0, and dot-dashed with slope $0.25$.}\label{Figure_9}
\end{figure}

The negative region of $\partial_y\xztavg{U}$ implies a counterflow in the mid-channel of the meanflow. fig.~\ref{Figure_9} 
shows profiles of the meanflow, with and without the laminar contribution, for the rotation numbers near the narrow 
and broad peaks.
Most of the profiles have a monotonically increasing slope. In various high-$Re$ non-rotating experiments of 
plane Couette flow, Reichardt \cite{Reichardt1959} showed, and later supported theoretically by 
Busse \cite{busse1970bounds}, that the meanflow in the center of the channel has a slope of $+1/4$; this 
is arguably confirmed for the non-rotating case $(a)$ and those near the broad peak, $(e)$ and $(f)$. However, 
panels $(b)$ and $(c)$, which correspond to the narrow peak, show a nearly flat region near the middle of the channel where the profile displays a slight negative slope, and therefore a weak counterflow. For this nearly flat region in the total meanflow, $\xztavg{U}=\uwall y+\xztavg{u}$, the laminar flow is compensated by the mean of the deviation term, $\xztavg{u}$, which from the previous figure, it is known that the laminar flow is slightly over-compensated.

There are some additional features of the $\xztavg{u}$-profile to be noted. Firstly, there is a turning-point by each wall corresponding to high-speed fluid near the wall being advected towards the opposite wall.
Despite a lack of continually coherent vortices, the non-rotating flow still produces this profile, suggesting that loosely and intermittently cooperating vortices have an impact. The peaks of $\xztavg{u}$ change non-monotonically with $\Ros$, reaching a maximum between the narrow and broad peaks.  
The second feature of the $\xztavg{u}$-profiles is that there is a `width' associated with the peaks. 
Considering first for larger rotation rates, a strong crossflow quickly sweeps the streamwise flow to the opposing walls, deforming the profile there rather than in the center; this results in the narrowing of the $\xztavg{u}$-profiles' peaks. In contrast, at low-$\Ros$, weaker vortices move the streamwise component through the mid-channel more slowly, which promotes coupling between the streamwise and cross-flows and results in broader maxima.

\begin{figure}[h]
\includegraphics[scale=0.5]{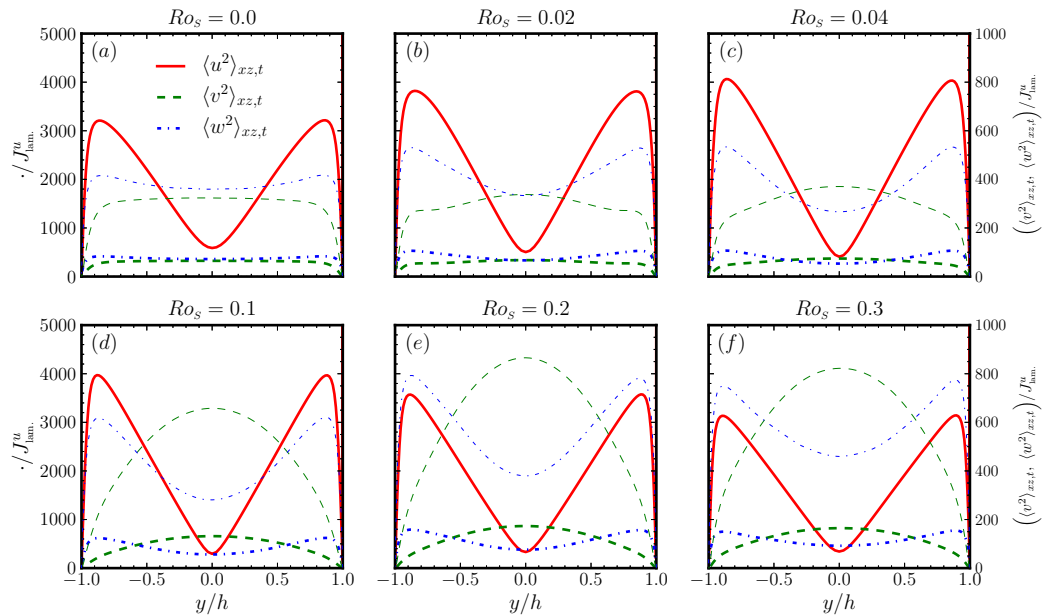}
\caption{The quasi-Reynolds stress profiles for the normal components, $\xztavg{u^2_i}$, at $Re =  5200$, using rotation numbers associated with the narrow (top row) and broad (bottom row) peaks. The normalization is made with the laminar momentum current, ${J}^u_{\scriptscriptstyle{\text{lam.}}}$. Note, the thin-lined representations of the crossflow components scale with the right ordinate axis.}\label{ProfilesRMS}
\end{figure}

Since the cooperation of the streamwise and crossflow velocities is believed to be responsible for the narrow peak via the quasi-Reynolds stress in the momentum current, it is reasonable to consider the mean velocities in the mid-channel where $\xztavg{uv}$ is largest. However, the crossflow means vanish and in order to get a measure of the their magnitudes, we instead observed the quasi-Reynolds stress profiles of the squared velocity components, $\xztavg{u^2_i}$, defined using the same spatial-averaging in Eq.~\ref{define_xztavg} with additional time-averaging. These are plotted in fig.~\ref{ProfilesRMS}, where to compare against the momentum current, they are normalized with ${J}^u_{\scriptscriptstyle{\text{lam}}}$. Note that these stresses also provide information from fluctuations about the means using the relation 
$\Xztavg{(\Delta u_i)^2} = \Xztavg{(\xztavg{u_i}-u_i)^2}$; 
in the case of the crossflow, the stresses are equal to the squared fluctuations whereas this only holds for $u$ at the mid-channel where the mean streamwise flow vanishes.

Firstly, for all graphs, these quantities are significantly larger than $Nu$. Secondly, in all graphs, the streamwise stress is  dominant in most, if not all, parts of the channel; 
it is where it is not dominant that seems relevant. In graphs $(a)$ and $(b)$, the mid-channel minimum of $\uu$, and hence $\xztavg{(\Delta u)^2}$, is still stronger than the crossflow stresses/fluctuations. 
Without rotation, the spanwise component in the mid-channel is stronger than with rotation; 
this is consistent with the observed flowfields, as shown in fig.~\ref{ProfilesRMS}$(a)$. 
At anti-cyclonic rotation just above zero, the baseflow destabilizes, large coherent vortices form, 
and the wall-normal flow develops. 
fig.~\ref{ProfilesRMS}$(b)$ then shows that at $\Nux$ the mid-channel regions of $\vv$ and $\ww$ intersect, with $\vv$ continuing to increase until just after $\Ros\sim0.04$, where it matches $\uu$. Thus, when the flow transitions towards a quasi-$2D$ state, $\vv$ is the largest in the mid-channel, as shown in fig.~\ref{ProfilesRMS}$(c)$.

Following the increase in $\Ros$, both $\vv$ and $\ww$ continuously increase with the strengthening of 
the vortices until the broad maximum, fig.~\ref{ProfilesRMS}$(e)$. Already by $\Ros=0.1$,  fig.~\ref{ProfilesRMS}$(d)$, 
$\vv$ is quite large in the mid-channel, taking a parabolic shape that is different from its profile for lower $\Ros$; its maximum is larger than those of $\ww$. At $\Ros=0.2$, both $\vv$ and $\ww$ have increased further and in the middle of the channel $\ww$ coincides with $\uu$. For even larger rotation rates, both components of the crossflow are stronger than the streamwise flow in the mid-channel.

Returning to the profiles, we offer the interpretation that the strength of the streamwise flow, which affects both the quasi-Reynolds stress component and the gradient of the mean flow, is largest in the center of the channel until the force reaches its maximum.
It is also significant that  $\vv$ surpasses $\ww$ here since this denotes an increasing wallwise flow in this region, allowing for a better coupling with the streamwise flow. The mid-channel strengths of $\uu$ and $\vv$ are decreasing and increasing, respectively, and they intersect for $\Ros>0.04$. This alludes to a narrow range of $\Ros$ where there is an optimal mixing that causes the slight streamwise counterflow in the mid-channel, as in fig.~\ref{Figure_9}$(b)$. 

\section{Conclusions}
In this study, we have numerically explored the turbulent plane Couette system with anti--cyclonic global rotation imposed. This study was carried out for moderate-to-large Reynolds numbers, $Re=650,\,1300,\,2600,\,5200$. In accordance with other numerical results for RPC \cite{bech1995investigation,bech1996secondary,bech1997turbulent,barri2010computer}, and similar to recent experiments in Taylor-Couette systems \cite{PhysRevLett_106_024501,PhysRevLett_106_024502}, our results show that the momentum transport $Nu$ is strongly influenced by the rotation. However, as we have demonstrated, changes in the momentum flux with rotation are attributed to different flow states and do not come from explicit rotation related terms in the
momentum balance.

For all Reynolds numbers, the turbulent momentum flux can be mainly attributed to the underlying Taylor vortex solution
over a wide range for $\Ros\gtrsim 0.04$. This finding was supported through comparisons between $Nu$ calculated from exact solutions of the TVF and from the turbulent simulations for $Re = 1300$, as well as examinations of the velocity fields and their spectral representations. 

One of the more intriguing findings reported here is a marked deviation from this TVF-behavior, where for $Re \gtrsim 2600$ and within a narrow range of low rotation numbers, $\Ros=0.01-0.04$, there is an abrupt increase in $Nu$, seen as a second peak. It is also significant that for $2600 < Re < 5200$, this narrow peak becomes larger in amplitude than the broad peak associated with the TVF, creating a discontinuous shift in the $\Ros$-position of the flux maximum. The flow associated with this narrow peak consists of a single pair of somewhat flattened counter-rotating vortices and includes a streamwise modulation, resembling the wavy vortex flow also identified in this and the TC-system. This modulation is seen in the streamwise Fourier modes, $\widetilde{E}_{l,0}$, which is apparent in all Reynolds numbers.

Analysis of the mean-velocity profiles and quasi-Reynolds stresses shows distinct relationships among the velocity fluctuations, described here using $\xztavg{(\Delta u_i)^2}$, associated with the peaks; this highlights the crucial role the coupling between the streamwise and wallwise velocities plays in the transport maxima. In the broad peak, the mid-channel fluctuations of the wallwise velocity are the strongest followed by the streamwise fluctuations; this relationship is reversed when the narrow peak emerges for large $Re$. The coincidence, and role, of the downstream modulation found in the narrow peak's flow-state remains an open issue.

The results here are obtained in boxes which are small compared to other studies mentioned in an earlier sections. We noted agreement between our results and those from those large domain simulations, though we should stress that the dynamics are constrained by Eq.~\ref{mmtm_current_Nu}, which is independent of the length and width of the boxes. Non-periodic boundary
conditions in the spanwise direction will change the flux balance, but one can expect that for sufficiently
wide domains the properties from the periodically continued ones can be recovered. 

The observation that the 2-d simulations can capture much of the torque of the 3-d systems may be viewed as an 
extreme example of a coherent structure (in this case the 2-d vortices) dominating the properties of a turbulent flow.
It would be interesting to trace this state and follow its bifurcations as parameters are changed, and to check
where and to which extend it continues to dominate the momentum flux. 

\section*{Acknowledgements}
We would like to thank Hannes Brauckmann for many fruitful discussions and comments.
This work was supported by the Deutsche Forschungsgemeinschaft (DFG).

\bibliography{SalewskiEckhardt_B}
\bibliographystyle{unsrt}

\end{document}